\documentclass[]{spie}  

\usepackage{amsmath,amsfonts,amssymb}
\usepackage{graphicx}
\usepackage[colorlinks=true, allcolors=blue]{hyperref}

\title{Deep learning enables extraction of capillary-level angiograms from single OCT volume}

\author[a]{Jianlong Yang}
\author[b]{Peng Liu}
\author[a]{Yan Hu}
\author[b]{Lixin Duan}
\author[c,a]{Jiang Liu}
\affil[a]{Cixi Institute of Biomedical Engineering, Chinese Academy of Sciences, China}
\affil[b]{Big Data Research Center at University of Electronic Science and Technology of China}
\affil[c]{Department of Computer Science and Engineering, Southern University of Science and Technology, China}
\authorinfo{Send correspondence to J.Y.: E-mail: yangjianlong@nimte.ac.cn}

\begin{document} 
\maketitle

\begin{abstract}
Optical coherence tomography angiography (OCTA) has drawn numerous attentions in ophthalmology. However, its data acquisition is time-consuming, because it is based on temporal-decorrelation principle thus requires multiple repeated volumetric OCT scans. In this paper, we developed a deep learning algorithm by combining a fovea attention mechanism with a residual neural network, which is able to extract capillary-level angiograms directly from a single OCT scan. The segmentation results of the inner limiting membrane and outer plexiform layers and the central $1\times1$ mm$^2$ field of view of the fovea are employed in the fovea attention mechanism. So the influences of large retinal vessels and choroidal vasculature on the extraction of capillaries can be minimized during the training of the network. The results demonstrate that the proposed algorithm has the capacity to better-visualizing capillaries around the foveal avascular zone than the existing work using a U-Net architecture.
\end{abstract}

\keywords{Optical coherence tomography angiography, deep learning, ocular imaging}

\section{INTRODUCTION}
Optical coherence tomography angiography (OCTA) is an emerging functional imaging modality of OCT. It is capable of discriminating blood flow of retinal vessels with surrounding static tissue without the dye injection process, which is obligatory in traditional Fluorescein Angiography (FA). Besides, its capacity in resolving depth vascular information has shown superiority in the early detection and diagnosis of prevailing ocular diseases, such as diabetic retinopathy , age-related macular degeneration, and glaucoma \cite{gao16}.

However, existing OCTA techniques require much longer acquisition time than OCT, which brings difficulties to the eye fixation of the observed subject. OCTA requires to continuously scan multiple times at the same region to calculate the 3D angiogram, because it follows the physical principle that the temporal decorrelation of the back-scattered photons is related to its velocity and the time interval bewteen two measurements \cite{zhu17}. Numerous technical efforts have been spent to eliminate or minimize motion-induced noise or artifacts \cite{camino16}, but the increased cost and complexity of the OCTA systems hinder the clinical popularization of this promising new imaging modality. \\
\indent Machine learning especially deep learning has the capability of extracting high-dimensional information, which implies that these techniques may be able to locate the retinal vessels and capillaries without the requirement of multiple repeated scans. In 2018, Lee \textit{et al.} firstly tested this possibility with an U-Net architecture \cite{lee18}. The generated cross-sectional and \textit{en face} angiograms looks promising. However, the visualization of the capillaries around fovea avascular zone (FAZ) and the contrast of the vessels still have room for improvement. \\
\indent In this paper, we propose a new deep-learning-based algorithm to extract capillary-level angiograms from single volumetric retinal OCT data. A fovea-attention mechanism using the segmentation information of the central fovea region is employed to mask out the large retinal vessel and the choroidal regions, which is beneficial for the model to learn better representations of the capillaries. 
\section{METHODS AND MATERIALS}
\begin{figure}[htb]
    \centering
    \includegraphics[width=12cm]{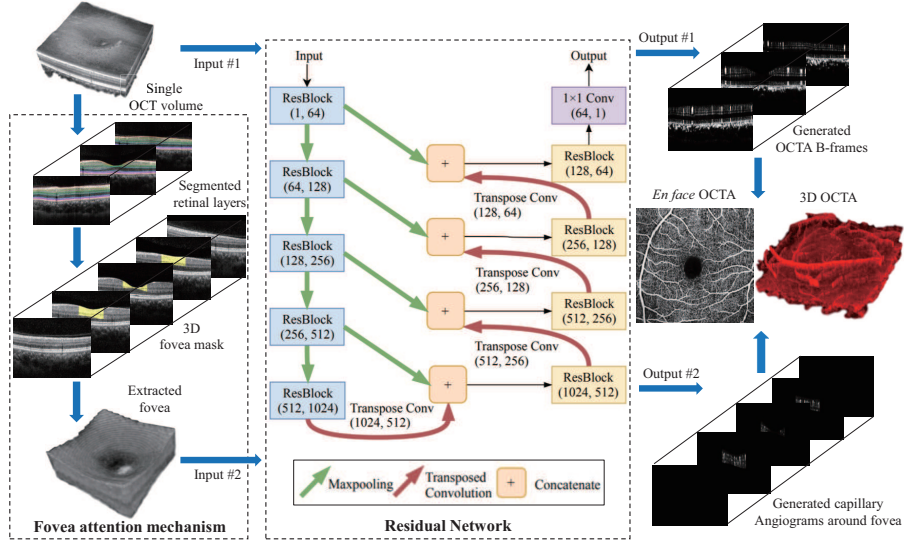}
    \caption{Workflow of the proposed algorithm.}
    \label{fig1}
\end{figure}

Fig. \ref{fig1} illustrates the workflow of the proposed algorithm. To extract the full retinal flow maps and the capillary angiograms around the fovea, the full B-frames and the B-frames masked with the fovea-attention mechanism are input to the residual U-Net separately. The residual U-Net firstly extracts the features from the B-frames and then generates the retinal flow maps by upsampling the features gradually. The outputs of this network are the retinal flow map of full OCTA B-frames and the capillary angiograms around the fovea. Then they are combined to generate the \textit{en face} and 3D OCTA. The fovea mask, from the \textit{en face} view, is a square containing fovea and a part of tissue surrounding the FAZ. From the B-frame perspective, its upper boundary is the inner limiting membraneand and the lower boundary is the outer plexiform layers.\\
\indent The deep learning network consists of downsampling and upsampling residual learning \cite{he2016deep}, which employs the shortcut mechanism to avoid the gradient vanishing and accelerate the network convergence. To mitigate the influence of information loss caused by downsampling, we apply the skip connections \cite{ronneberger2015u} to concatenate the downsampled features and upsampled features on the same level. \\
\indent We established an OCTA dataset using a spectral-domain OCT system with OCTA module (SPECTRALIS, Heidelberg Engineering, Germany) to evaluate our model. It contains $36$ OCT volumes from both eyes of 18 healthy volunteers. Each volume has 384 B-frames. They are grayscale images with $496\times 379$ pixels. \textbf{Each OCT B-frame has the corresponding decorrelation-based OCTA images as the ground truth.} We take $3$ volumes as the test set and take the other $33$ volumes as train and validation set. 
\section{RESULTS AND DISCUSSION}
\begin{figure}[htb]
    \centering
    \includegraphics[width=14cm]{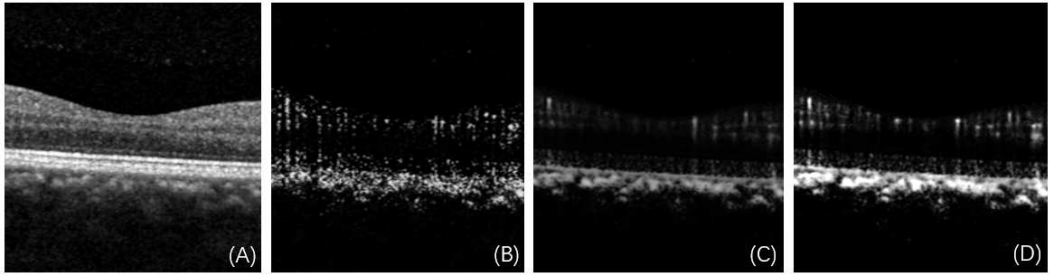}
    \caption{(A) OCT B-frame. (B) ground truth. (C) U-Net-generated OCTA B-frame. (D) this work. }
    \label{fig2}
\end{figure}
Fig. \ref{fig2} is the comparison of the generated OCTA B-frames using the deep learning algorithms [(C) is the U-Net baseline and (D) is this work] with the original OCT (A) and the decorrelation-based OCTA ground truth (B). The capillaries are barely seen in the OCT B-frame but can be extracted by both the decorrelation-based and deep-learning-based OCTA algorithms. The proposed fovea-attention network demonstrates better capability of extracting the capillaries close to the foveal center than the U-Net baseline.\\ 
\begin{figure}[htb]
    \centering
    \includegraphics[width=14cm]{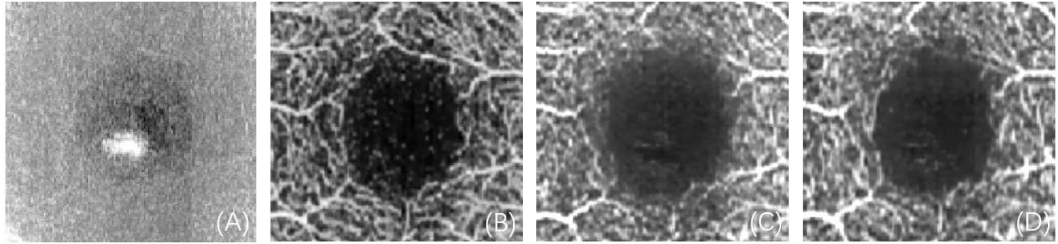}
    \caption{(A) \textit{En face} OCT. (B) ground truth. (C) U-Net-generated \textit{En face} OCTA. (D) this work.}
    \label{fig3}
\end{figure}
\indent \textit{En face} OCTA is frequently used in ophthalmology because the quantitative biomarkers like vessel density and flow index can be extract from it conveniently. Fig. \ref{fig3} demonstrates the result comparison in \textit{en face} view. The visibility and contrast of the capillaries around the FAZ are better in the \textit{en face} OCTA image using the proposed algorithm than the U-Net baseline. Also, the shape and area of the FAZ in Fig.~3(D) are very similar to the decorrelation-based ground truth in Fig.~3(B). 
\section{CONCLUSIONS}
The proposed method is promising in complementing the existing temporal-decorrelation-based OCTA methods because it avoids the requirement of the complex OCTA hardware and/or software, thus benefiting the popularization of this state-of-the-art imaging functionality in the study and diagnosis of ocular diseases.
\acknowledgments 
This work is supported by Ningbo 3315 Innovation team grant and Cixi Institute of Biomedical Engineering, Chinese Academy of Sciences (Y60001RA01, Y80002RA01).

\bibliography{OCTA} 
\bibliographystyle{spiebib} 

\end{document}